\documentclass[preprintnumbers,amsmath,amssymb,showpacs,twocolumn]{revtex4}

\usepackage{graphicx}
\usepackage{dcolumn}
\usepackage{bm}

\begin{document}

\title{ Three-party simultaneous quantum secure direct communication scheme with EPR pairs}

\author{Mei-Yu Wang, Feng-Li Yan}
\thanks {Corresponding author. Email address: flyan@mail.hebtu.edu.cn}
\affiliation {College of Physics  and Information Engineering, Hebei
Normal University, Shijiazhuang 050016,
China\\
}

\date{\today}

\begin{abstract}

  We present a scheme for three-party simultaneous quantum secure direct communication by using EPR pairs. In the scheme,
   three legitimate parties can simultaneously exchange their secret messages. It is also proved to be
   secure against the intercept-and-resend attack, the disturbance attack and the entangled-and-measure attack.

  \end{abstract}


\maketitle

\section{Introduction}

Quantum key distribution (QKD) has been one of the most promising
applications of quantum mechanics, which provides unconditionally
secure key exchange. The basic principle is that two remote
legitimate users (Alice and Bob) establish a shared secret key
through the transmission of quantum signals and use this key to
encrypt or decrypt the secret messages. Since Bennett and Brassard
\cite {s1} proposed the standard QKD protocol in 1984 (BB84), a good
many QKD protocols have been advanced \cite {s2, s3, s4, s5, s6}.

 In recent years, a new concept, quantum secure direct communication
 (QSDC) has been proposed \cite {s7, s8, s9, s10, s11,s12,s13}.
 Different from quantum key distribution whose object is to
 establish a common random key between two parties, quantum secure direct communication
 is to transmit the secret messages directly without first
 establishing a random key to encrypt them.  Bostr$\ddot{\mathrm
 o}$m and Felbinger's protocol \cite {s8}, which was called  the
 ping-pong protocol,  allows decoding the encoded bits
 instantaneously and directly in each respective transmission run.
 However, it was proved to be insecure against the disturbance
 attack \cite {s14}. Deng et al \cite {s9} have proposed a two-step
 quantum direct communication protocol by using EPR pairs. In this
 two-step scheme, the EPR pairs are divided into two sequences,
 which are sent by two steps, and the receiver needs to check the
 security of the channel twice (one for checking sequence and
 another for message sequence). In these schemes, QSDC is only one
 way communication. Based on the idea of ping-pong QSDC scheme,
 Nguyen \cite {s15} proposed a quantum dialogue scheme by using EPR
 pairs that enables both legitimate parties to exchange their secret
 messages in a direct way. However, it is not secure against the
 intercept-and-resend attack. An eavesdropper who adopts this attack
 strategy can steal the secret messages without being detected. More
 recently, Jin et al \cite {s16} proposed a three-party
 simultaneous QSDC scheme by using the GHZ states.

 In this paper, we will introduce a three-party (Alice, Bob and
 Charlie) simultaneous QSDC scheme by using EPR pairs. In our
 scheme, each can obtain two other's messages simultaneously.
 Moreover, it will be shown that this protocol is provably secure.

\section{The three-party simultaneous QSDC scheme with EPR pairs}

Suppose that Alice, Bob and Charlie have a secret message
respectively, without loss of generality we assume that the message
length of three parties is the same.
\begin{eqnarray}
&&\mathrm{Alice's ~~~message}=\{i_1,i_2,i_3,\cdots,i_N\},\\
&&\mathrm{Bob's~~~ message}=\{j_1,j_2,j_3,\cdots,j_N\},\\
&&\mathrm{Charlie's~~~ message}=\{k_1,k_2,k_3,\cdots,k_N\},
\end{eqnarray}
with $i_n,j_n,k_n\in \{0,1\}.$

Alice, Bob and Charlie agree on that Bob  performs the two unitary
operations
\begin{equation}
C_{j_n}=\{\begin{array}{lll} I, & \mathrm{if} &
j_n=0,\\
\sigma_x, & \mathrm{if} & j_n=1,\\
\end{array}
\end{equation}
Charlie operates the two unitary operations
\begin{equation}
C'_{k_n}=\{\begin{array}{lll} I, & \mathrm{if}
&k_n=0,\\
\sigma_z, & \mathrm{if}& k_n=1,\\
\end{array}
\end{equation}
where
\begin{equation}
I=|0\rangle\langle 0|+|1\rangle\langle 1|, \sigma_x=|0\rangle\langle
1|+|1\rangle\langle 0|, \sigma_z=|0\rangle\langle
0|-|1\rangle\langle 1|.
\end{equation}

To secretly exchange their messages, Alice first produces a large
enough number of entangled pairs all in the state
\begin{equation}
|\Psi_{00}\rangle_{ht}=\frac {1}{\sqrt
2}(|0\rangle|1\rangle+|1\rangle|0\rangle)_{ht}.
\end{equation}
Here $h$ and $t$ denote home particle remaining in the place of
Alice, and transmitting particle  being transmitted in the
communication, respectively. Then Alice, Bob and Charlie proceed as
follows:

(S0) Protocol is initialized $n=0$.

(S1) Set  $n=n+1$.  Alice keeps qubit $h_n$ with her and sends qubit
$t_n$ to Bob through the quantum channel.

(S2) When Bob received a qubit, he has two choices: one is  to
measure it, the other is  to encode it.

2.1 If Bob decides to measure the qubit, that means to check the
eavesdropping, he can  complete it by the following procedure:

(a) Bob chooses randomly one of the two sets of measuring basis
(MB), say $\{|0\rangle, |1\rangle\}$ and $\{|+\rangle, |-\rangle\},$
to measure the qubit $t_n$.

(b) Bob tells Alice the MB  chosen by him and the outcome of his
measurement.

(c) Alice chooses the same MB to measure the qubit $h_n$ and checks
with the result of Bob. If no eavesdropper  exists their results
should be correlated, i.e., if Alice gets $|0\rangle$ ($|1\rangle$),
Bob will get $|1\rangle$ ($|0\rangle$), when they choose their
measurements along the $Z$-direction; or if Alice gets $|+\rangle$,
($|-\rangle$), Bob will obtain $|+\rangle$ ($|-\rangle$) when they
choose the MB $\{|+\rangle, |-\rangle\}$. If their results are
correlated, set $n=n-1$, go to step 1 and continue communication.
Otherwise, they abort communication.

2.2 If Bob wants to encode his messages, he selects a running mode
from message mode (MM)and control mode (CM).

1) If he selects MM, he encodes $j_n$ by performing the
transformation $C_{j_n}$ on the qubit $t_n$.

2) If he selects CM, he does nothing.

Then Bob sends the qubit $t_n$ to Charlie.

(S3) Charlie confirms Bob that he received a qubit. Then Bob
announces the running mode.

3.1 If it was run in CM, obviously the qubits $t_n$ and $h_n$ are in
the state $|\Psi_{00}\rangle_{ht}$, Charlie and Alice check
eavesdropper as procedure 2.1.

3.2 If it was run in MM, Charlie selects a running mode from his two
modes: MM and CM.

1) If he selects MM, Charlie encodes $k_n$ by performing the
transformation $C'_{k_n}$ on the qubit $t_n$, and then he sends it
to Alice.

2) If he selects CM, Charlie randomly prepares a decoy qubit in one
of states $\{|0\rangle, |1\rangle, |+\rangle, |-\rangle\}$ and sends
it to Alice.

(S4) Alice confirms Charlie that she received a qubit. Then Charlie
announces the running mode.

4.1) If the running mode was CM, Charlie publicly reveals the qubit
state. Alice measures the decoy qubit using the corresponding basis.
If the result is in accordance with Charlie's result, set $n=n-1$
and go to step 1, otherwise abort the communication.

4.2) If the running mode was MM, Alice measures $t_n$ and $h_n$ in
the Bell basis. Suppose the result is
$|\Psi_{r_ns_n}\rangle_{h_nt_n}$, Alice encodes her message into
$r_n, s_n$ by
\begin{equation}
r_n\oplus i_n=x_n, ~~ s_n\oplus i_n=y_n.
\end{equation}
Then Alice announces publicly the values of $(x_n, y_n)$.

If $n<N$, go to step 1.

(S5) The three-party simultaneous QSDC has been successfully
completed.

Now, we explicitly analyze the protocol described above. After Bob
encodes his bit $j_n$ on the EPR-pair state
$|\Psi_{00}\rangle_{h_nt_n}$, the pair state becomes
\begin{equation}
|\Psi_{00}\rangle_{h_nt_n}\rightarrow
C_{j_n}|\Psi_{00}\rangle_{h_nt_n}=|\Psi_{j_n0}\rangle_{h_nt_n}.
\end{equation}
Furthermore, after Charlie encodes his message, the state is
transformed as
\begin{equation}
|\Psi_{j_n0}\rangle_{h_nt_n} \rightarrow
C'_{k_n}|\Psi_{j_n0}\rangle_{h_nt_n}=|\Psi_{j_nk_n}\rangle_{h_nt_n},
\end{equation}
where
\begin{eqnarray}
&& |\Psi_{10}\rangle_{h_nt_n}=\frac {1}{\sqrt
2}(|0\rangle|0\rangle+|1\rangle|1\rangle)_{h_nt_n},\\
&& |\Psi_{01}\rangle_{h_nt_n}=\frac {1}{\sqrt
2}(|1\rangle|0\rangle-|0\rangle|1\rangle)_{h_nt_n},\\
&& |\Psi_{11}\rangle_{h_nt_n}=\frac {1}{\sqrt
2}(|0\rangle|0\rangle-|1\rangle|1\rangle)_{h_nt_n}.
\end{eqnarray}
Therefore if no eavesdropper exists, Alice's measurement result will
be
\begin{equation}
r_n=j_n, s_n=k_n.
\end{equation}
Clearly, if Alice publicly announces the values of $r_n, s_n$, the
messages of Bob and Charlie will be revealed. For realizing the
secure QSDC, Alice plays a trick: she encodes her message into $r_n$
and $s_n$, which was described in Eq. (8).

By Eqs.(8) and (14), we conclude easily that
\begin{equation}
j_n\oplus k_n=x_n \oplus y_n.
\end{equation}
Obviously, Alice, Bob and Charlie can obtain messages of other
parties by Eqs.(8), (14) and (15). Decoding rules can be expressed
as
\begin{eqnarray}
&& \mathrm{Alice} ~~~ j_n=x_n\oplus i_n, ~~ k_n=y_n\oplus i_n,\\
&& \mathrm{Bob}~~~
i_n=x_n\oplus j_n,  ~~ k_n=x_n\oplus y_n\oplus  j_n,\\
 && \mathrm{Charlie} ~~~ i_n=y_n\oplus
k_n,  ~~j_n=x_n\oplus y_n\oplus k_n.
 \end{eqnarray}
From the above decoding rules we may find that the messages $i_n,
j_n$ and $k_n$ play two-fold role in the QSDC: on the one hand, they
represent the messages of three parties; on the other hand, they
play a role of the keys of three parties. However, it is worth
pointing out that these keys are different from a private key which
is prior shared by the parties of communication, but three parties
encode these keys to the EPR pairs, so these keys hide in the
entanglement.

\section{Security of QSDC}

In the following, we briefly discuss the security of the scheme.

Firstly, we consider the intercept-and-resend attack and disturbance
attack. In both two attacks, Eve would destroy the entanglement of
particles $h_n$ and $t_n$ in $A\rightarrow B$ line or $B \rightarrow
C$ line. In $C \rightarrow A$ line, Eve's attack will cause
disaccord between Charlie's decoy qubit state and Alice's
measurement outcome. Each of the above cases will make Eve be
detcted with  a detection probability of 1/2. Thus our scheme is
secure against the intercept-and-resend attack and disturbance
attack.

Eve's another attack is entangle-and-measure attack \cite {s9}. It
acts in the following way. Eve prepares an ancilla in the initial
state $|\chi\rangle_e$ and waits in the route. When the qubit $t_n$
passes by, Eve entangles his ancilla with it by  performing a
unitary operation $\hat {E}$ defined as
\begin{eqnarray}
&&\hat
{E}|0\rangle_t|\chi\rangle_e=\alpha|0\rangle_t|\chi_0\rangle_e+\beta|1\rangle_t|\chi_1\rangle_e,\\
&&\hat
{E}|1\rangle_t|\chi\rangle_e=\alpha|1\rangle_t|\chi_0\rangle_e+\beta|0\rangle_t|\chi_1\rangle_e,
\end{eqnarray}
with $\alpha,\beta$ satisfying the normalization condition
$|\alpha|^2+|\beta|^2=1$ and $\{|\chi_0\rangle_e,
|\chi_1\rangle_e\}$ being the pure orthonormalized ancilla's states
uniquely determined by the unitary operation $\hat E$. Obviously,
Eve's attack in $A\rightarrow B$ or $B\rightarrow C$ destroys the
correlation of EPR pairs. The state of EPR pair and the ancilla
becomes
\begin{equation}
\hat {E}|\Psi_{00}\rangle_{ht}|\chi\rangle_e=
\alpha|\Psi_{00}\rangle_{ht}|\chi_0\rangle_e+\beta|\Psi_{10}\rangle_{ht}|\chi_1\rangle_e,
\end{equation}
Eve conceals himself if his measurement ends up with
$|\chi_0\rangle_e$. However if the measurement outcome is
$|\chi_1\rangle_e$, he will be detected. If Eve's attack is in the
$C \rightarrow A$ line, besides the above Eqs. (19) and  (20), the
attack on the decoy qubit in $|+\rangle$ $(|-\rangle)$ is expressed
as
\begin{eqnarray}
&&\hat
{E}|+\rangle_t|\chi\rangle_e=|+\rangle_t(\alpha|\chi_0\rangle_e+\beta|\chi_1\rangle_e),\\
&&\hat
{E}|-\rangle_t|\chi\rangle_e=|-\rangle_t(\alpha|\chi_0\rangle_e-\beta|\chi_1\rangle_e),
\end{eqnarray}
In this case, the eavesdropper can not be detected. However the
eavesdropper can be found by Charlie and Alice when the decoy qubit
is in the state $|0\rangle$ or $|1\rangle$. So our scheme is also
secure against this attack.

\section {Conclusion}

In summary, we have proposed a protocol for  three-party quantum
secure direct communication by using EPR states. Our scheme has its
own advantages compared with the schemes proposed previously.
Firstly, three legitimate parties can simultaneously exchange their
secret messages. Secondly, compared with  the preparation of GHZ
states, obviously the EPR state will be more easily prepared in the
experiment.

 \acknowledgments  This work was supported by the National  Natural Science Foundation of
China under Grant No: 10671054 and Hebei Natural Science Foundation
of China under Grant No: A2005000140.

\end{document}